# Cell nucleus elastography with the adjoint-based inverse solver

Yue Mei[1,2,3], Xuan Feng[1,2], Yun Jin[1,2], Rongyao Kang[1,2], XinYu Wang[1,2], Dongmei Zhao[1,2], Soham Ghosh[4], Corey P. Neu[5,6,7], Stephane Avril[8,*]

[1]State Key Laboratory of Structural Analysis, Optimization and CAE Software for Industrial Equipment, Department of Engineering Mechanics, Dalian University of Technology, Dalian, 116023, China

[2]International Research Center for Computational Mechanics, Dalian University of Technology, Dalian 116023, P.R. China

[3]Ningbo Institute of Dalian University of Technology, No. 26 Yucai Road, Jiangbei District, Ningbo, 315016, China

[4]Department of Mechanical Engineering, Colorado State University, Fort Collins, CO, USA

[5]Paul M. Rady Department of Mechanical Engineering, University of Colorado Boulder, Boulder, CO, USA

[6]Biomedical Engineering Program, University of Colorado Boulder, Boulder, CO, USA

[7]BioFrontiers Institute, University of Colorado Boulder, Boulder, CO, USA

[8]Mines Saint-Étienne, Univ Jean Monnet, INSERM, U 1059 Sainbiose, F - 42023, Saint-Étienne, France

Corresponding author*: avril@emse.fr

**Abstract**

**Background and Objectives:** The mechanics of the nucleus depends on cellular structures and architecture, and impact a number of diseases. Nuclear mechanics is yet rather complex due to heterogeneous distribution of dense heterochromatin and loose euchromatin domains, giving rise to spatially variable stiffness properties.

**Methods:** In this study, we propose to use the adjoint-based inverse solver to identify for the first time the nonhomogeneous elastic property distribution of the nucleus. Inputs of the inverse solver are deformation fields measured with microscopic imaging in contracting cardiomyocytes.

**Results:** The feasibility of the proposed method is first demonstrated using simulated data. Results indicate accurate identification of the assumed heterochromatin region, with a maximum relative error of less than 5%. We also investigate the influence of unknown Poisson's ratio on the reconstruction and find that variations of the Poisson's ratio in the range [0.3-0.5] result in uncertainties of less than 15% in the identified stiffness. Finally, we apply the inverse solver on actual deformation fields acquired within the nuclei of two cardiomyocytes. The obtained results are in good agreement with the density maps obtained from microscopy images.

**Conclusions:** Overall, the proposed approach shows great potential for nuclear elastography, with promising value for emerging fields of mechanobiology and mechanogenetics.

# 1. Introduction

The study of cellular and nuclear mechanics has gained increasing attention in recent years due to its essential role in the regulation of gene expressions [1-2] affecting cellular function such as cell migration [3-5], proliferation or differentiation [6-7]. A number of diseases have been associated to altered cellular mechanics, as for instance, atherosclerosis and other cardiovascular diseases [8]. Analysis of full-field strain maps depicting the behavior of cardiomyocyte nuclei during in vitro contractions has revealed a discernible relationship between nuclear deformation, chromatin reorganization, and the subsequent reduction in the expression of genes associated with cardiac development [9]. Consequently, there is a pressing need in elucidating the intricate mechanosensitive role of the cell nucleus in the pathophysiology of cardiovascular diseases. To name another example, osteoarthritis [10-11], which is a common joint disorder affecting millions of individuals worldwide, it is now well understood that changes in the extracellular matrix, cell and nuclear mechanics are associated with the cartilage degeneration [12]. Many other pathological conditions can be attributed to compromised mechanobiological pathways that are triggered by or result into compromised cell and nuclear mechanics [13-15].

To investigate cell mechanics, a variety of methods have been developed to measure their elastic and viscoelastic properties [16]. Among these methods, atomic force microscopy (AFM) is commonly used to estimate the nano-indentation response of cells and relate it to their viscoelastic properties [17-18]. However, the indentation response combines mechanical contributions of subcellular structures including the cell membrane and nucleus which are difficult to separate [16]. To better measure the separate mechanical contributions of each cellular compartment, particle tracking microrheology (PTM) was also developed [19-20], although applications to identify local elastic properties are rare.

A recent progress in the field is the introduction of optical elastography to measure full-field deformation fields induced by subjecting the cell to adapted mechanical loading [21]. By leveraging the theory of elastic waves [22] and assuming local homogeneity [23-24], the spatial variation of cell elasticity could be obtained. However, these elastic maps did not resolve spatial heterogeneities within the nucleus, whereas internal architecture of the cell nucleus show heterogeneous distribution of dense heterochromatin and loose euchromatin domains [25-27], normally giving rise to spatially variable stiffness properties [28]. Therefore, continued efforts to refine these techniques and develop new ones to measure the nonhomogeneous nucleus elasticity are essential.

Recent research has focused on characterizing the mechanical properties of euchromatin and heterochromatin due to their significant role in gene expressions [29]. Deformation microscopy, a non-invasive method based on high-resolution optical microscopy, has been proposed to obtain mechanical properties of the cell nucleus. This technique overcomes the challenges of AFM indentation as it does not require to invasively probe inside the cell nucleus [30]. While displacement and strain fields could be acquired successfully by this method, attempts to solve the inverse problem for relating them to local elastic properties has been limited by technical challenges. Only a recent study used an optimization scheme to estimate the effective elastic moduli of the heterochromatin and euchromatin regions based on pixel grayscale intensity [28], but this method reduced the inverse problem to the identification of only two stiffness parameters and neglected the possible heterogeneity within heterochromatin domains. There still remains a need to develop a complete description of intranuclear mechanics, including detailed spatial distributions of material properties, that may give insights into cellular structure, architecture, and biological processes.

In this paper, we address this technically challenging problem with an adjoint-based inverse solver without any *a priori* assumption about the distribution of stiffness properties within the nucleus. The paper is structured as follows: In the **Methods** Section, we describe the experimental setup of deformation microscopy and the inverse solver algorithm. In the **Numerical Example** Section, we demonstrate the feasibility of the proposed method using simulated datasets generated by finite-element simulations. In the **Experimental Example** Section, we identify the nonhomogeneous distribution of nuclear stiffness using the full-field displacement fields measured by deformation microscopy in beating cardiomyocytes. Then we discuss and analyze the numerical and experimental results in the **Discussion** Section. Finally, we conclude the paper with a summary of our findings and suggestions for future work.

2.Methods

2.1 Deformation microscopy within the cell nucleus

To characterize the nonhomogeneous elastic property distribution of the nucleus, the first step is to measure the displacement field inside a deforming nucleus using deformation microscopy. As depicted in **Fig.1**, cardiomyocytes were cultured on a PDMS substrate with two different stiffness values resembling a normal (soft, E~12kPa) vs fibrotic (stiff, E~434kPa) heart. High-resolution images of the beating cardiomyocyte nucleus at the medial plane was captured using a epifluorescence microscope. The two-dimensional displacement fields in the imaging plane was quantified by deformation microscopy, a technique based upon digital image correlation and hyperelastic warping, that utilizes the undeformed and deformed images. Further details regarding the experimental arrangements and displacement calculation can be found in [30].

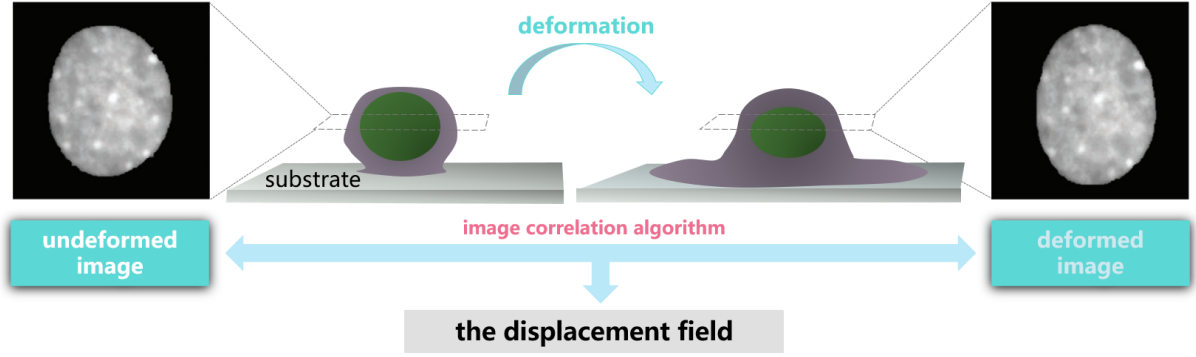

**Fig.1**: Schematic representation of the deformation microscopy setup to measure the displacement fields within the cell nucleus

## 2.2 Inverse problem solution algorithm

As the image of the cell nucleus is two-dimensional, the cell nucleus is assumed to be in a state of plane stress in its medial plane. It is also assumed that the nucleus behaves like an incompressible [31-32] and neo-Hookean material such as:

$$\mathbf{S} = \mu(\mathbf{I} - \mathbf{C}^{-1}) + pJ\mathbf{C}^{-1} \qquad (1)$$

where $\mathbf{S}$ is the second Piola-Kirchhoff stress tensor, $\mu$ is the shear modulus, $p$ denotes the hydrostatic pressure, $\mathbf{C} = \mathbf{F}^T\mathbf{F}$, $J = \det(\mathbf{F})$, where $\mathbf{F}$ is the deformation gradient. $\mathbf{F}$ derives from the displacement field $\mathbf{u}$ such as: $\mathbf{F} = \mathbf{I} + \mathbf{Grad}\ \mathbf{u}$. For a compressible and neo-Hookean material, the constitutive equation is given by:

$$\mathbf{S} = \mu(\mathbf{I} - \mathbf{C}^{-1}) + \frac{2\mu\nu}{1-2\nu}\ln(J)\mathbf{C}^{-1} \qquad (2)$$

where $\nu$ is the Poisson's ratio.

The inverse problem is stated as: with the measured displacement field $\mathbf{u}^m$ obtained by deformation microscopy, we seek the spatial distribution of shear modulus $\mu(x)$ within the nucleus that minimizes the following objective function:

$$\pi(\mathbf{u},\mu) = \frac{1}{2}\int_\Omega |\mathbf{u}-\mathbf{u}^m|^2 d\Omega + \alpha \int_\Omega \sqrt{|\nabla\mu|^2 + c^2}\, d\Omega \tag{3}$$

where $\Omega$ represents the entire domain of the nucleus and $\mathbf{u}$ is the displacement field satisfying quasi-static equilibrium such as div($\mathbf{FS}$)=0 in $\Omega$ and $\mathbf{u}=\mathbf{g}$ in $\Gamma$, $\Gamma$ being the boundary of $\Omega$. The shear modulus $\mu(x)$ is nodally defined and interpolated by the same shape function as the displacement fields. The second term in the objective function (regularization term) is used to avoid overfitting; $\alpha$ is the regularization factor that is used to control the significance of the regularization. For a very small $\alpha$, the elasticity map of the nucleus has strong oscillations and distortions, whereas it is very smooth if $\alpha$ is very large. In this paper, the optimal regularization factor is chosen by the L curve method [33]. $c=10^{-2}$ is a stabilizing factor mitigating potential singularity arising from the differentiation of the objective function with respect to the shear modulus.

We utilize the finite-element method to solve quasi-static equilibrium and compute $\mathbf{u}$. The weak form is defined as: Find $[\mathbf{u}, p] \in M \times P$ such that:

$$A(\mathbf{w},\mathbf{u},q,p) = 0, \quad \forall [\mathbf{w},q] \in \ell \times P \tag{4}$$

where

$$A(\mathbf{w},\mathbf{u},q,p) = \int_\Omega \nabla\mathbf{w}:\mathbf{P}d\Omega + \int_\Omega q(J-1)d\Omega \tag{5}$$

The function spaces $M$ and $\ell$ are defined as:

$$\mathsf{M} \equiv \{\mathbf{u} | u_i \in H^1(\Omega); u_i = g_i \text{ on } \Gamma_u\}$$
$$\ell \equiv \{\mathbf{w} | w_i \in H^1(\Omega); w_i = 0 \text{ on } \Gamma_u\} \quad (6)$$
$$\mathsf{P} \subseteq L_2(\Omega)$$

The domain of interest is discretized with quadrilateral elements and the corresponding discretized formulation is stated as:

Find $[\mathbf{u}^h, p^h] \in \mathsf{M}^h \times \mathsf{P}^h$ such that

$$A(\mathbf{w}^h, \mathbf{u}^h, q^h, p^h) = 0, \quad \forall [\mathbf{w}^h, q^h] \in \ell^h \times \mathsf{P}^h \quad (7)$$

where $\mathsf{M}^h$, $\ell^h$ and $\mathsf{P}^h$ are the finite-element subspaces of $\mathsf{M}$, $\ell$ and $\mathsf{P}$, respectively. Here, we use the standard Galerkin finite-element method. To address volumetric locking induced by the incompressibility assumption, a stabilization term is introduced into the weak form (7) such as:

$$A^*(\mathbf{w}^h, \mathbf{u}^h, q^h, p^h) = A(\mathbf{w}^h, \mathbf{u}^h, q^h, p^h) - \sum_{i=1}^{NE} \left(\tau \text{div}(\mathbf{F}^h \mathbf{S}^h), (\mathbf{F}^h)^{-T} \nabla q^h\right)_{\Omega^i} \quad (8)$$

where NE is the total number of finite elements across the discretized nucleus. Moreover, $\tau = \dfrac{\alpha h^2}{2\mu}$ represents the stabilization factor, where $h$ is the characteristic element length. $\alpha$ is set to 1.0. More details on the stabilization method can be found in [34].

To solve the inverse problem, we use the adjoint-based optimization method [35-37]. In this case, the Lagrangian is:

$$L(\mathbf{w}^h, \mathbf{u}^h, q^h, p^h; \mu) = \pi(\mathbf{u}^h, \mu) + A^*(\mathbf{w}^h, \mathbf{u}^h, q^h, p^h) \quad (9)$$

where $\mathbf{w}^h$ and $q^h$ can be interpreted as Lagrange multipliers. Note that if $A^*(\mathbf{w}^h, \mathbf{u}^h, q^h, p^h) = 0$, we obtain the following equality: $L = \pi$.

The variation of $L$ with respect to $\delta\mathbf{u}^h, \delta p^h, \delta\mathbf{w}^h, \delta q^h$ and $\delta\mu$ may be written such as

$$\delta L = D_{\mathbf{u}^h} L \, \delta\mathbf{u}^h + D_{p^h} L \, \delta p^h + D_{\mathbf{w}^h} L \, \delta\mathbf{w}^h + D_{q^h} L \, \delta q^h + D_\mu L \, \delta\mu \tag{10}$$

where $D$ is the directional derivative operator. Setting $D_{\mathbf{w}^h} L \, \delta\mathbf{w}^h + D_{q^h} L \, \delta q^h = 0$ yields the state equation:

$$A^*(\delta\mathbf{w}^h, \mathbf{u}^h, \delta q^h, p^h) = 0, \quad \forall [\delta\mathbf{w}^h, \delta q^h] \in \ell^h \times P^h \tag{11}$$

**Eq. 11** is the forward problem, whose resolution yields $\mathbf{u}^h$ and $p^h$.

Subsequentially, if we set $D_{\mathbf{u}^h} L \, \delta\mathbf{u}^h + D_{p^h} L \, \delta p^h = 0$, the adjoint equation can be obtained:

$$B(\mathbf{w}^h, q^h, \delta\mathbf{u}^h, \delta p^h; \mathbf{u}^h, p^h) = -(\delta\mathbf{u}^h, \mathbf{u}^h - \mathbf{u}^m)_\Omega, \quad \forall [\delta\mathbf{u}^h, \delta p^h] \in M^h \times P^h \tag{12}$$

where $B(\mathbf{w}^h, q^h, \delta\mathbf{u}^h, \delta p^h; \mathbf{u}^h, p^h) = \dfrac{d}{d\varepsilon} A^*(\mathbf{w}^h, q^h, \mathbf{u}^h + \varepsilon\delta\mathbf{u}^h, p^h + \varepsilon\delta p^h)\bigg|_{\varepsilon \to 0}$ is the linearization of $A^*$.

Eq. (12) is the adjoint equation and $\mathbf{w}^h$ can be obtained accordingly. With the acquired state variables $\mathbf{u}^h$, $p^h$ and the dual variables $\mathbf{w}^h$, $q^h$, the gradient of the objective function with respect to the shear modulus is deduced such as:

$$\delta L = D_\mu L \, \delta\mu = \alpha \int_\Omega \frac{\nabla\mu \cdot \nabla\delta\mu}{\sqrt{|\nabla\mu|^2 + c^2}} d\Omega + C(\mathbf{w}^h, q^h, \delta\mu; \mathbf{u}^h, p^h, \mu) \tag{13}$$

where $C\left(\mathbf{w}^h, q^h, \delta\mu; \mathbf{u}^h, p^h, \mu\right) = \dfrac{d}{d\varepsilon} A^*\left(\mathbf{w}^h, q^h, \mathbf{u}^h, p^h, \mu + \varepsilon\delta\mu\right)\bigg|_{\varepsilon \to 0}$.

With the objective function and spatial variation of the objective function with respect to the shear modulus, we can update the shear modulus distribution using the L-BFGS method [38]. This process is repeated until the objective function, or the $L_2$ norm of gradients of the objective function with respect to nodal shear moduli, become lower than the machine precision.

In summary, the algorithm of the inverse solver is shown in **Fig.2**.

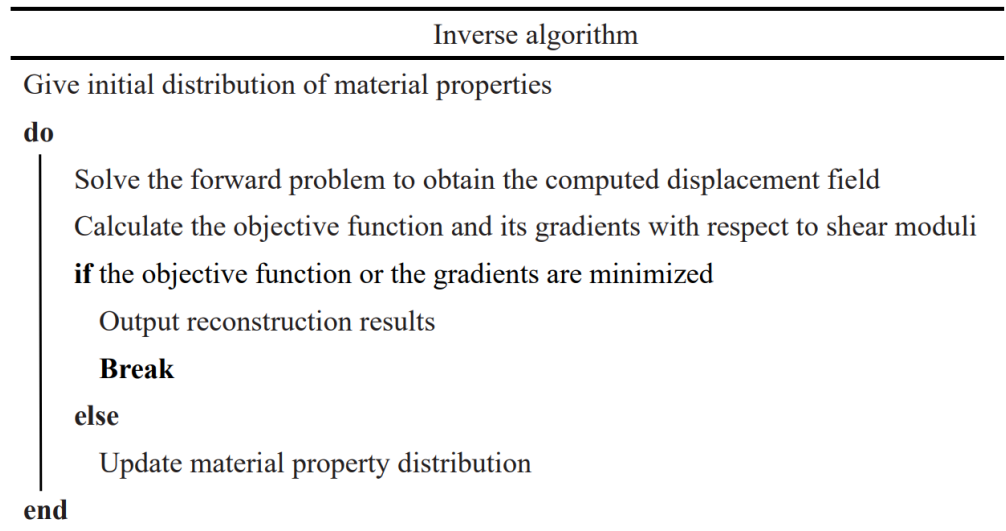

**Fig.2**: Inverse algorithm diagram

## 3. Results

### 3.1 Numerical Examples

The proposed inverse approach for the identification of nonhomogeneous shear modulus distribution in the cell nucleus was first tested with several simulated cases of deformation generated by finite element method (forward problem), as shown in **Fig.3**. We built a finite-element model of cell nucleus with regions of heterochromatin (H) and euchromatin (E) following the validation example introduced in [30], shown in **Fig.3**(b). Each domain was assigned a different stiffness, the A and B regions having the same stiffness because they belong to the same chromatin domain. Subsequently, the uniform tension of 500kPa along the outward normal of the boundary was applied to the boundary of the cell nucleus, and finite-element simulations were performed with the commercial FE software ABAQUS, while assuming an incompressible behavior. The resulting simulated full-field displacement fields were used as inputs for testing the inverse solver (then using simulated data instead of actual data). To account for the unknown force information in actual nuclear deformation microscopy. we have to apply the displacements on the entire boundary in the inverse problem. Furthermore, various levels of Gaussian noise were added to the simulated displacement fields to simulate measurement errors. The noise level was defined such as:

$$\text{noise level} = \sqrt{\frac{\sum_{i=1}^{NN}\left(u_i^m - \bar{u}_i\right)^2}{\sum_{i=1}^{NN}\left(\bar{u}_i\right)^2}} \times 100\% \tag{14}$$

where NN is the total number of nodes in the meshed nucleus, $\bar{u}_i$ and $u_i^m$ are the exact nodal displacement and the "noisy" nodal displacement at the *i*-th node, respectively.

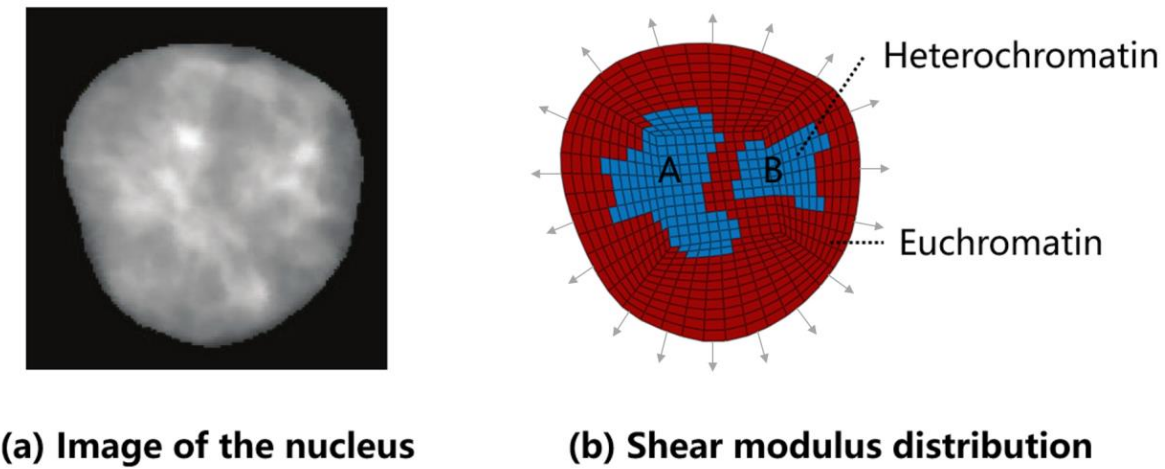

**Fig.3**: (a) Microscopy image of the nucleus in the medial plane; (b) Finite-element mesh and nonhomogeneous shear modulus distribution in the modeled nucleus.

The ratio between the shear modulus of the H (heterochromatin) region and the shear modulus of the E (euchromatin) region is denoted as the H:E ratio. This ratio is assigned in the finite-element model to simulate the nuclear displacement fields and the goal of the inverse solver is to recover the correct ratio, namely the target H:E ratio.

When the H:E ratio is 2:1, as depicted in **Fig.4**, the inverse solver is able to correctly recover the target shear modulus distribution even when the noise level is at 10%. However, as the noise level increases, errors increase at the boundaries of the H domain. In **Fig.5**, the displacement and strain fields from the forward simulation with different noise levels and the inverse results were thoroughly compared. We observed that with the increase of the noise level, the strain fields from the forward simulation become seriously noisy compared to the associated displacement fields. A similar trend can be observed in **Figs.6** and **7**, where the target H:E ratio is 10:1. However, the shape of the H domain deteriorates when the noise level exceeds 4%. As shown in **Table 1**, the average shear moduli in the two regions (A and B) of the H domain are very close to the target values. The maximum error is less than 5%, indicating a strong reliability of the inverse solver. These results demonstrate the feasibility and robustness of the inversion method for characterizing nonhomogeneous shear modulus distributions of the cell nucleus in the presence of a Gaussian noise.

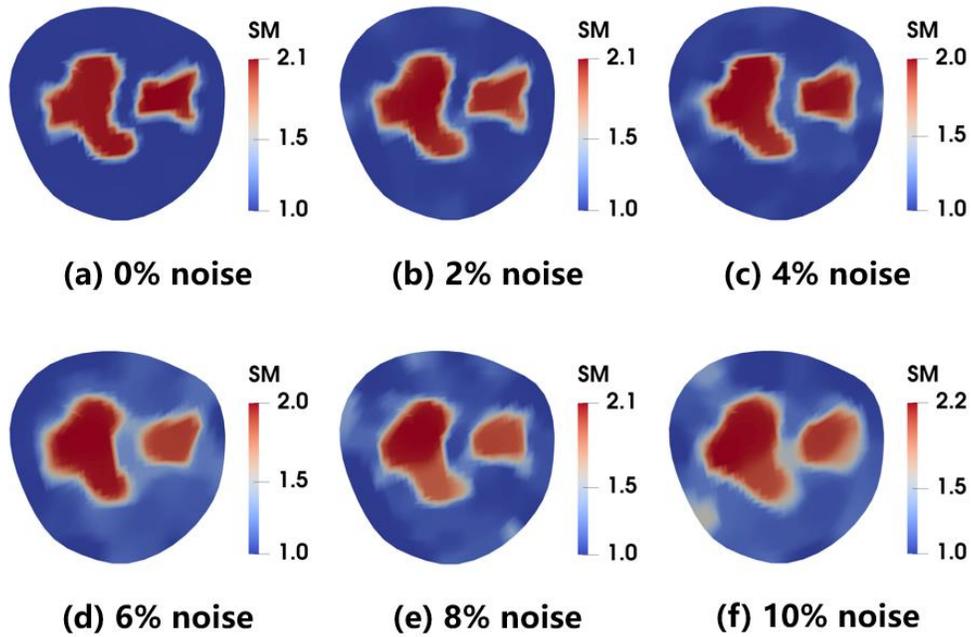

**Fig.4**: The reconstructed shear modulus distribution of the nucleus for varying noise levels when the target H:E ratio is 2:1

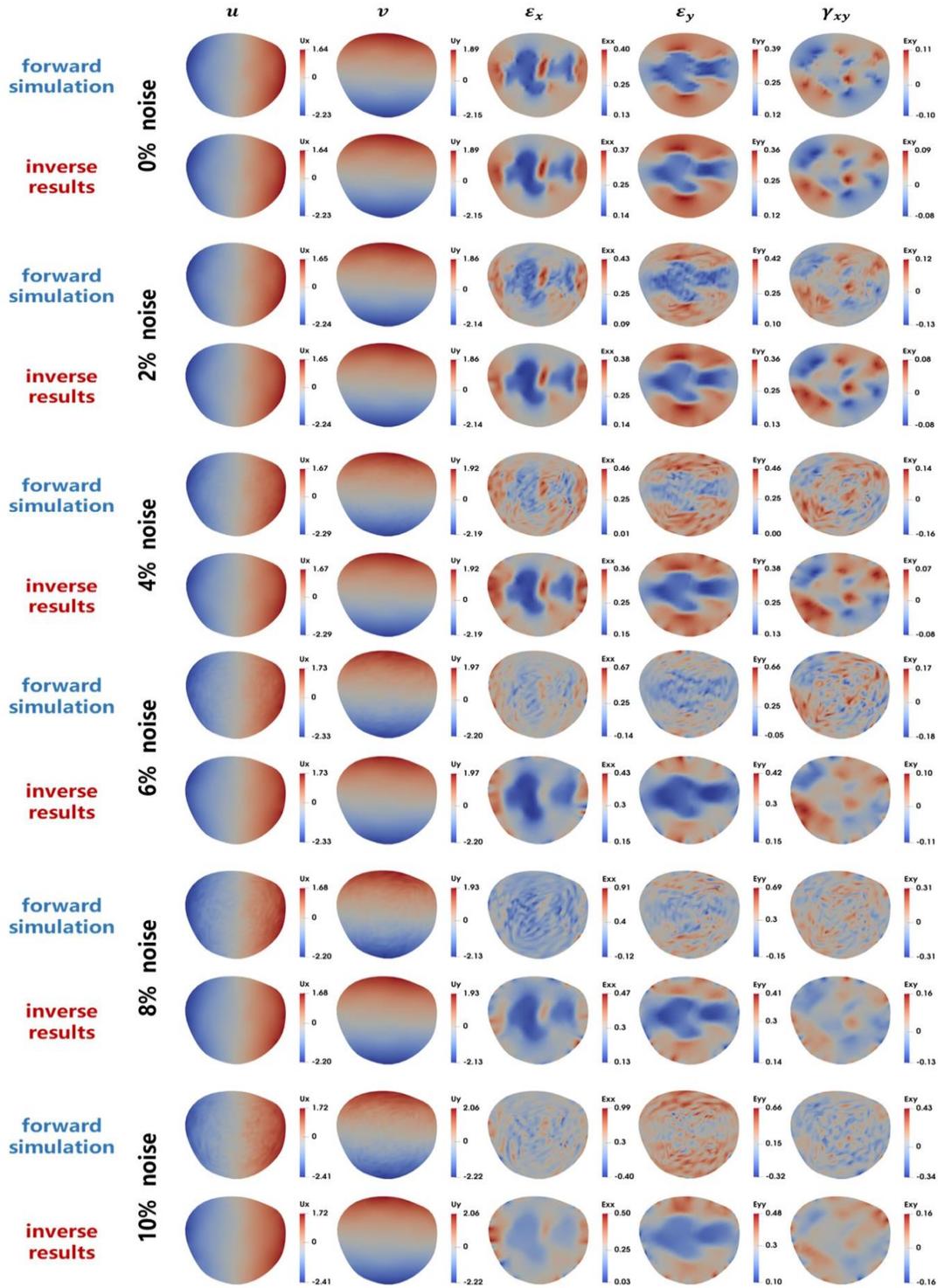

**Fig.5**: Displacement and strain fields obtained with the forward simulation and with the inverse results for all cases shown in **Fig.4**. The 1st and 2nd columns correspond to the x and y

displacement components, respectively. The 3$^{rd}$, 4$^{th}$, 5$^{th}$ columns correspond to three in-plane strain components.

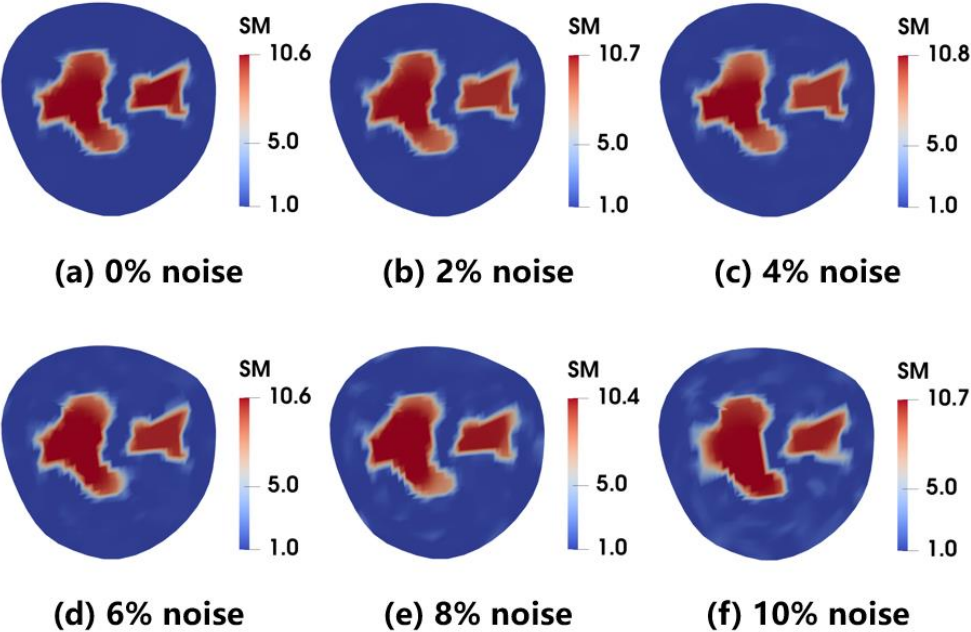

**Fig.6**: The reconstructed shear modulus distribution of the nucleus for varying noise levels when the target H:E ratio is 10:1

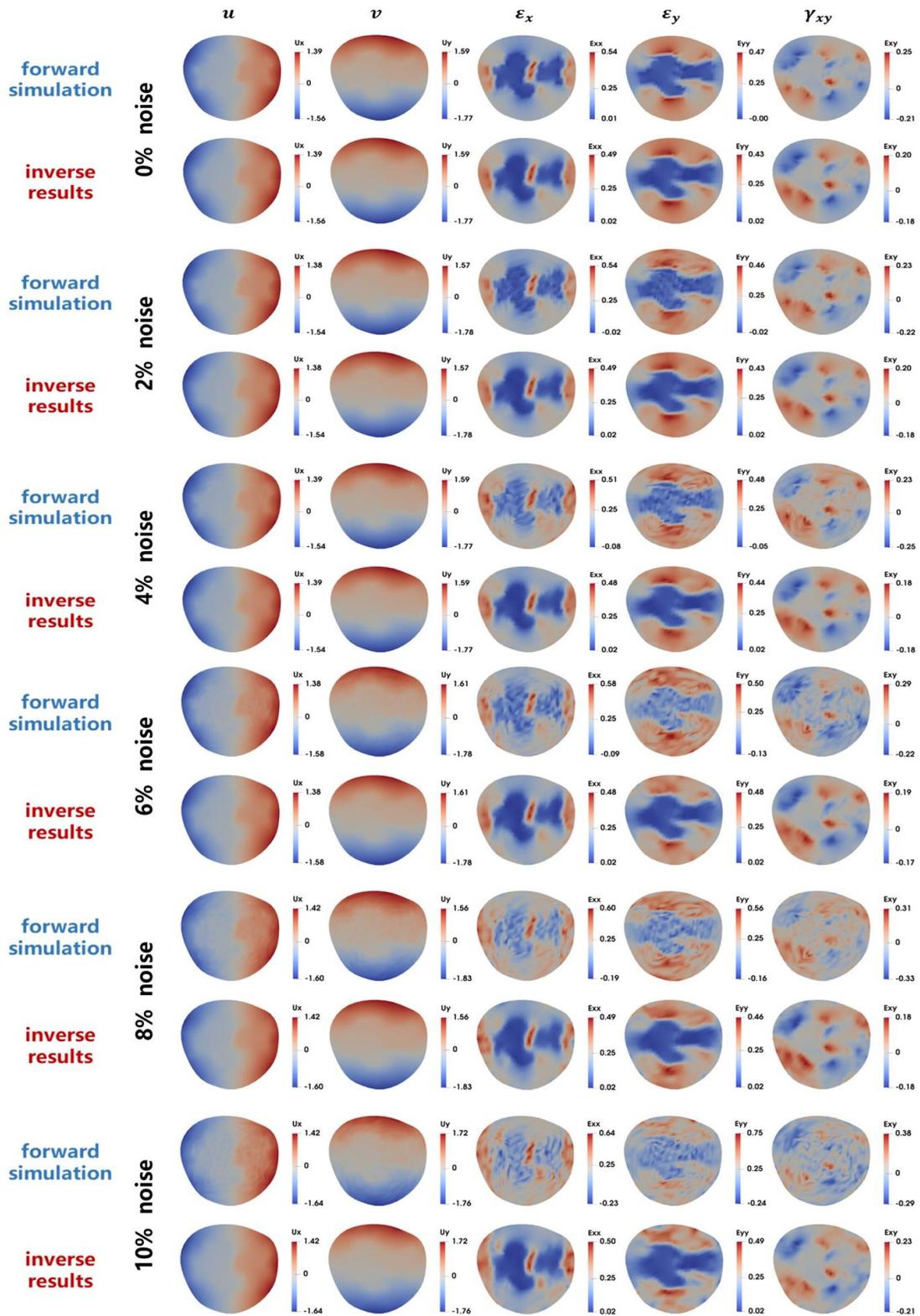

**Fig.7**: Displacement and strain fields obtained with the forward simulation and with the inverse results for all cases shown in **Fig.6.** The 1$^{st}$ and 2$^{nd}$ columns correspond to the x and y displacement components, respectively. The 3$^{rd}$, 4$^{th}$, 5$^{th}$ columns correspond to three in-plane strain components.

**Table 1**: Identified H:E ratios in regions A and B for the reconstruction shown in **Figs.4** and **6**

| Noise level/% | H:E=2:1 | | H:E=10:1 | |
| --- | --- | --- | --- | --- |
| | A | B | A | B |
| 0 | 2.03 | 2.05 | 10.06 | 10.29 |
| 2 | 2.02 | 2.00 | 10.06 | 9.99 |
| 4 | 2.00 | 1.98 | 9.99 | 10.03 |
| 6 | 2.01 | 1.90 | 10.13 | 10.08 |
| 8 | 2.04 | 1.94 | 9.84 | 9.92 |
| 10 | 2.09 | 2.00 | 10.34 | 9.81 |

Since the cell nucleus might be compressible [39], the influence of Poisson's ratio on the reconstructed results was considered. To this end, we simulated nuclear displacement fields with Poisson's ratio $v$=0.3 and checked what would be the impact on the inverse solver if we assume that the cell nucleus is incompressible in the inverse problem (fix the Poisson's ratio $v$=0.5 in solving the inverse problem). The obtained results demonstrate that the regions of the H domain could still be well recovered even when a wrong Poisson's ratio is assumed (see **Figs.8** and **10**). We also compared the displacement and strain fields from the forward simulation with different noise levels and the inverse results are shown in **Figs. 9** and **11**. Although relative error for H:E ratio is larger as shown from the calculated values, the maximum relative error in this case remains below 15% even in the presence of 10% noise (see **Table 2**). This demonstrates that the sensitivity of stiffness reconstruction to the Poisson's ratio is relatively small.

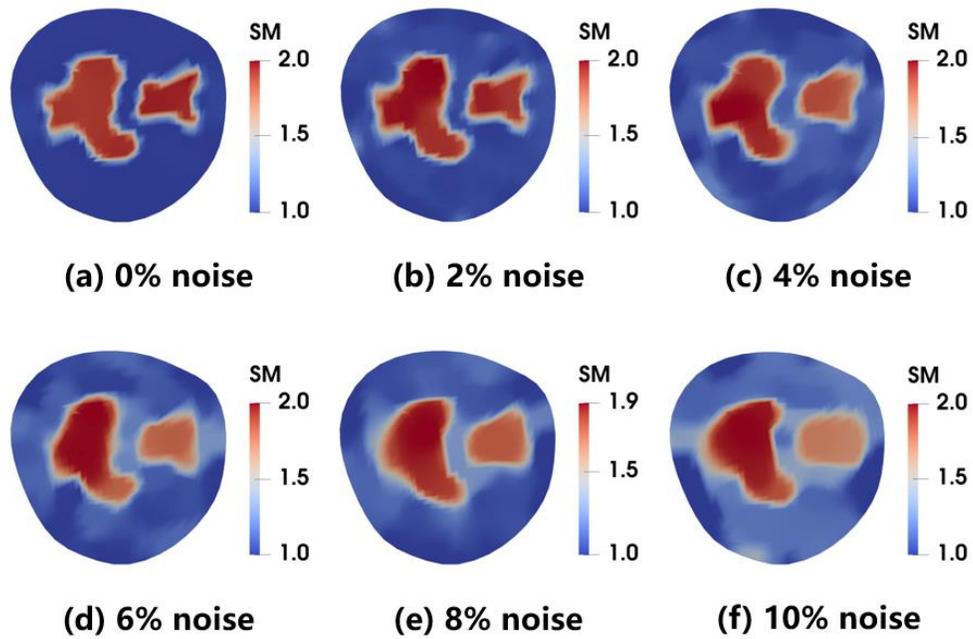

**Fig.8**: Reconstructed shear modulus distribution of the nucleus for varying noise levels when the target H:E ratio is 2:1 and when data are simulated with a finite-element model assuming the nucleus as compressible ($v$=0.3)

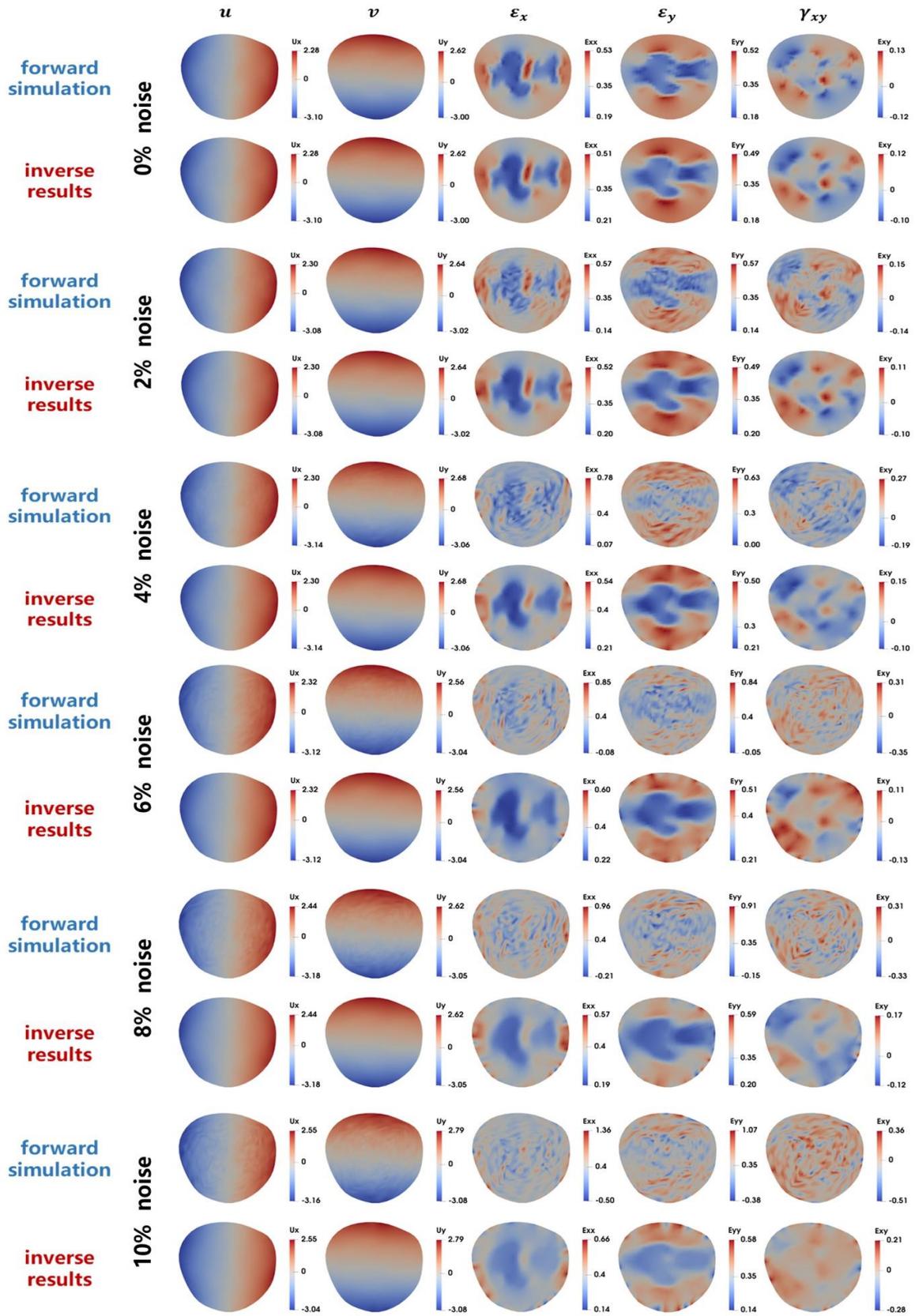

**Fig. 9**: Displacement and strain fields obtained with the forward simulation and with the inverse results for all cases shown in **Fig. 8.** The 1st and 2nd columns correspond to the x and y displacement components, respectively. The 3rd, 4th, 5th columns correspond to three in-plane strain components.

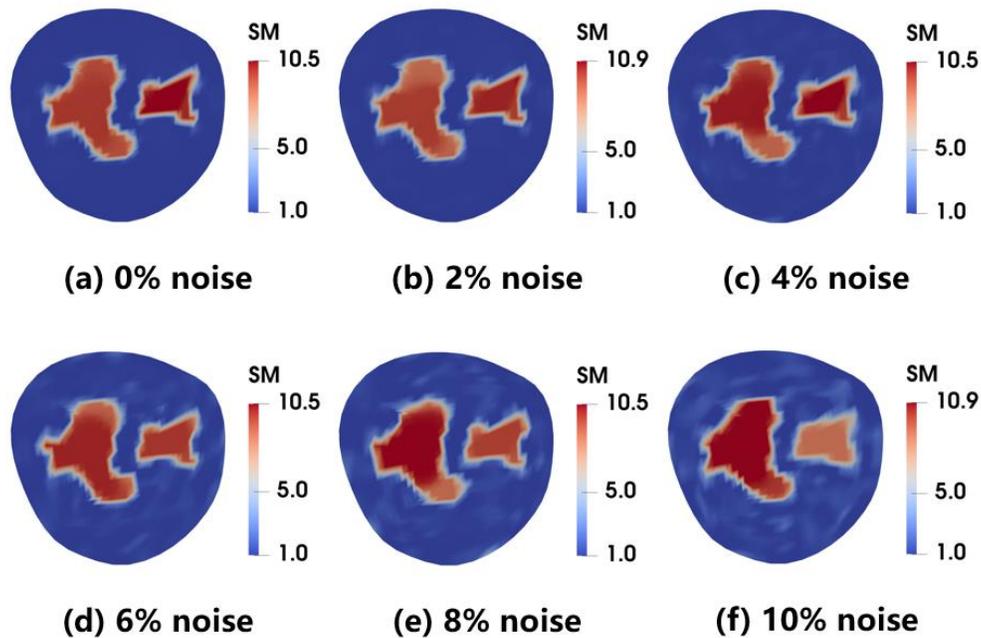

**Fig.10**: Reconstructed shear modulus distribution of the nucleus for varying noise levels when the target H:E ratio is 10:1 and when data are simulated with a finite-element model assuming the nucleus as compressible (*v*=0.3)

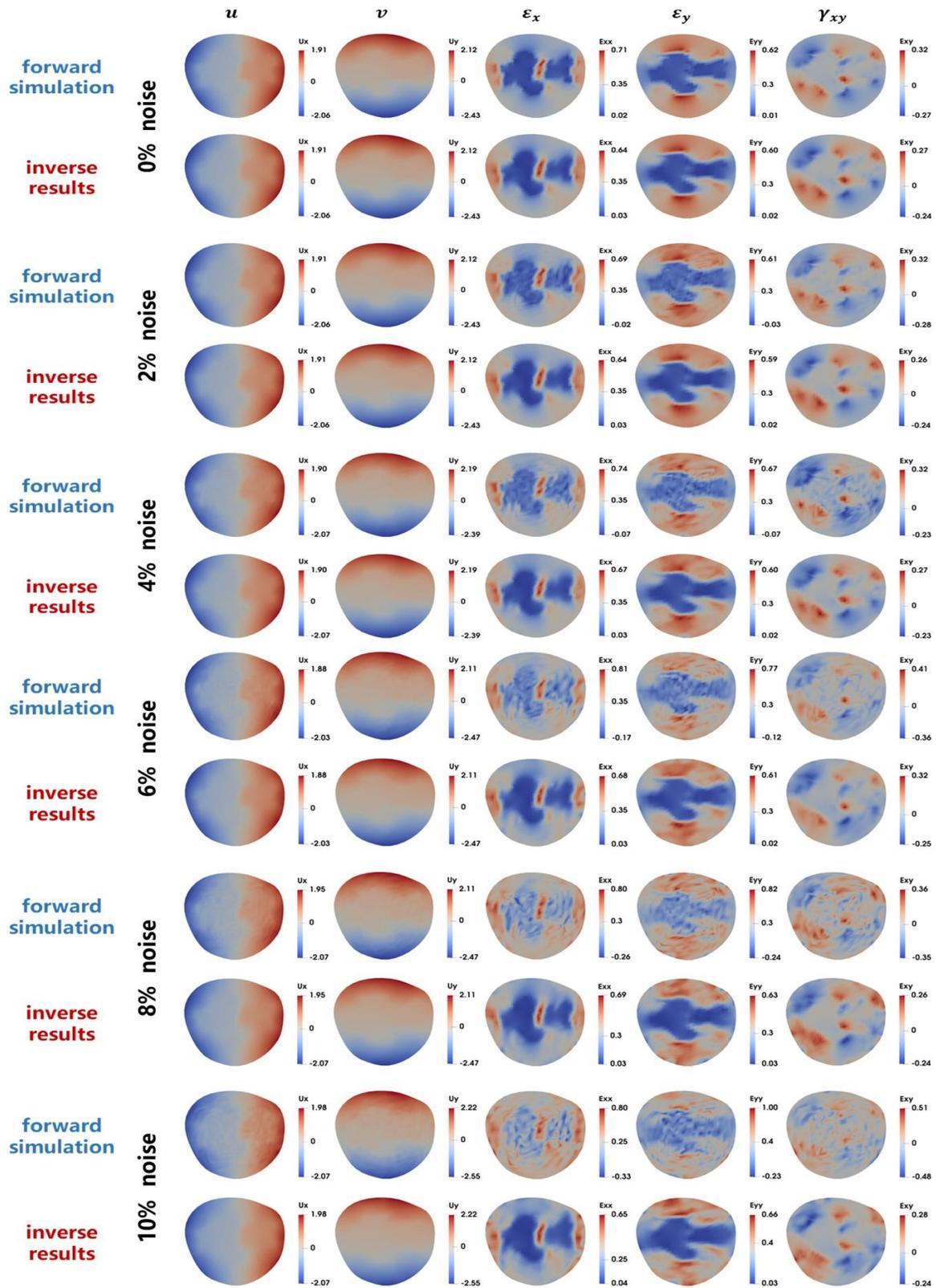

**Fig.11**: Displacement and strain fields obtained with the forward simulation and with the inverse results for all cases shown in **Fig. 10.** The 1$^{st}$ and 2$^{nd}$ columns correspond to the x and y displacement components, respectively. The 3$^{rd}$, 4$^{th}$, 5$^{th}$ columns correspond to three in-plane strain components.

**Table 2**: Identified H:E ratios in regions A and B for the reconstruction shown in **Figs.8** and **10**

| Noise level/% | H:E=2:1 | | H:E=10:1 | |
| --- | --- | --- | --- | --- |
| | A | B | A | B |
| 0 | 1.91 | 1.95 | 9.47 | 10.29 |
| 2 | 1.95 | 1.93 | 9.54 | 10.32 |
| 4 | 1.95 | 1.85 | 9.77 | 10.15 |
| 6 | 1.94 | 1.80 | 9.76 | 9.68 |
| 8 | 1.86 | 1.75 | 9.91 | 9.46 |
| 10 | 1.94 | 1.72 | 10.66 | 8.74 |

The reliability of the plane stress assumption necessitates validation. Therefore, we modeled a homogeneous, incompressible 3D spherical nucleus subjected to a uniform tension of 500kPa applied on its equatorial plane, as illustrated in **Fig.12(a)**. There are two ellipsoids with the same dimension embedded in the sphere. The sphere was discretized into 62084 tetrahedral elements. Using the displacement fields in the equatorial plane, we identified the elastic map with our inverse solver, as shown in **Fig.12(b)**. We assumed that the plane was under a state of plane stress. The reconstructed stiffness, as depicted in **Figs.12(c)** and **12(d)**, reveal that the H:E distributions align closely with the target. The relative error for H:E ratios is kept under 1% in both scenarios. In the strain fields shown in **Fig.13**, the two inclusions can be clearly observed from the strain image. These findings suggest that the bias due to the plane stress assumption in the identification process is not significant for a spherical nucleus.

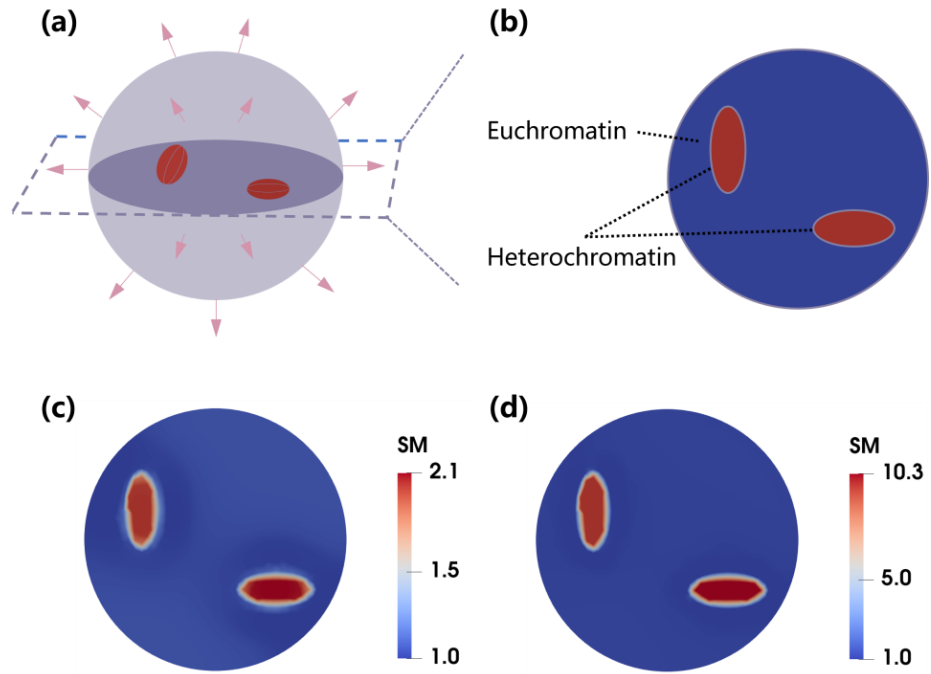

**Fig.12**: (a) 3D spherical nucleus subjected to a uniform tension in its equatorial plane; (b) Target stiffness map in the equatorial plane; (c) Identified stiffness distribution in the equatorial plane when the target H:E ratio is 2:1; (d) Identified stiffness distribution in the midplane when the target H:E ratio is 10:1.

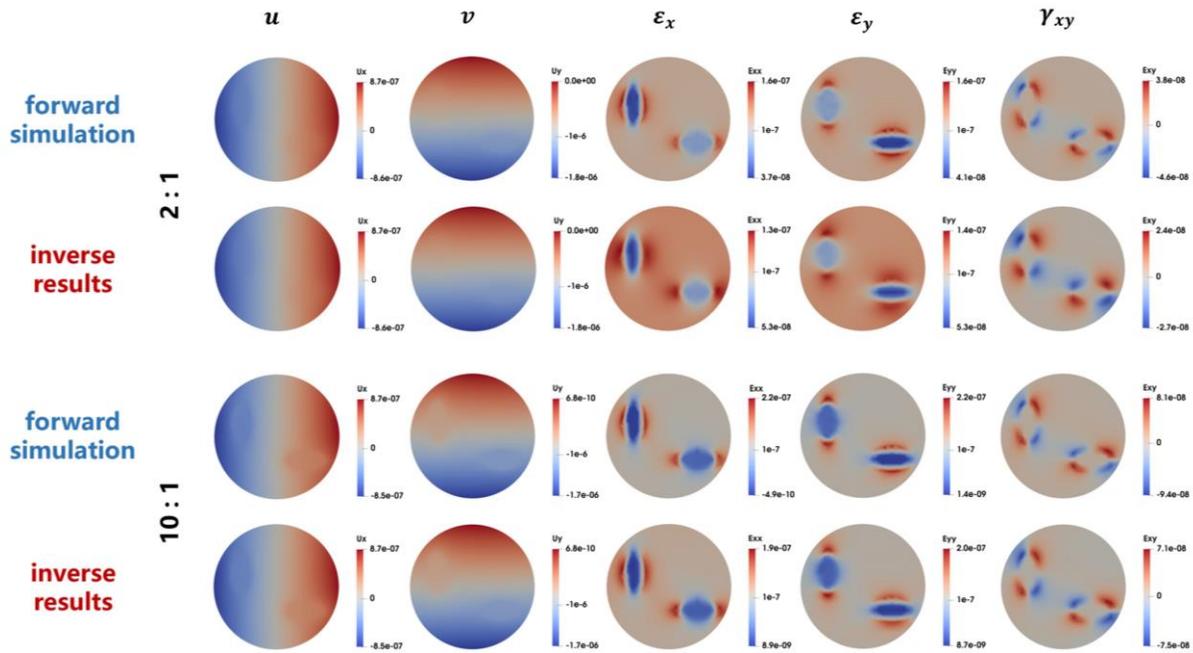

**Fig. 13**: Displacement and strain fields obtained with the forward simulation and with the inverse results for all cases shown in **Fig. 12.** The 1st and 2nd columns correspond to the x and y displacement components, respectively. The 3rd, 4th, 5th columns correspond to three in-plane strain components.

However, it is worth noting that the previous 3D model exhibited a significant degree of symmetry. We also explored an alternate 3D model configuration with two irregular inclusions embedded within the upper portion of the spherical nucleus, as shown in **Fig. 14**. For the mechanical loading conditions, we applied compression to the upper surface while constraining the lower region. To solve the inverse problem, we selected the plane corresponding to the upper surfaces of these two inclusions for analysis. When dealing with a lower target H:E (Hardness-to-Elasticity) ratio of 2:1, the quality of the reconstruction results was notably suboptimal. For a more substantial target H:E ratio of 10:1, although the two inclusions can be detected (as shown in the strain maps of **Fig.15**), we observed that the reconstructed H:E ratio deviated significantly towards 4.0, a value substantially lower than the target. This discrepancy highlights that the plane stress assumption is a very strong assumption in our study which is not satisfied in case on non-symmetric internal morphology of the cell nucleus.

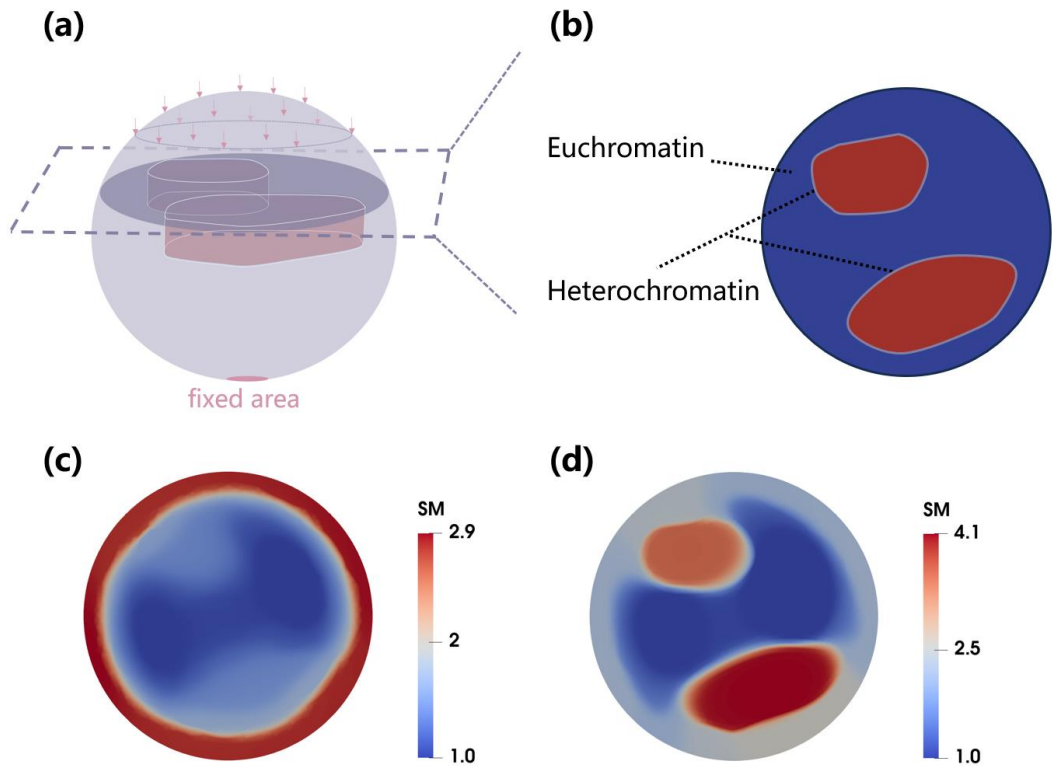

**Fig.14**: (a) 3D spherical nucleus subjected to a uniform compression on the upper surface with a non symmetric internal morphology; (b) Target stiffness map in the plane of interest; (c) Identified stiffness distribution in the equatorial plane when the target H:E ratio is 2:1; (d) Identified stiffness distribution in the midplane when the target H:E ratio is 10:1.

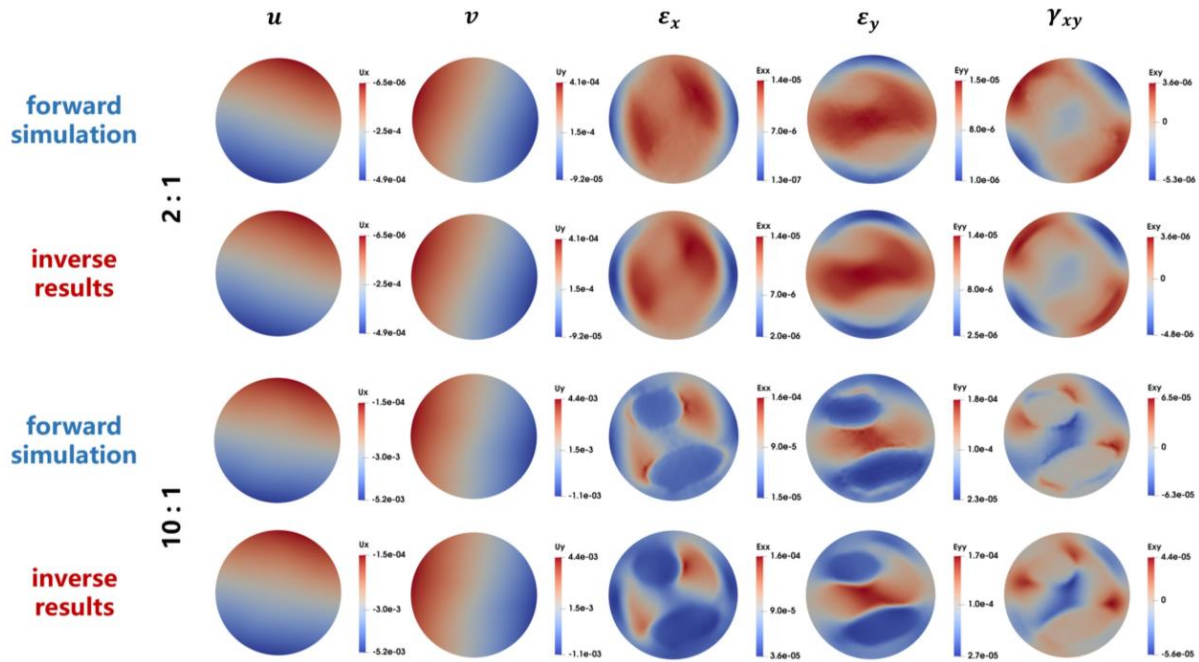

**Fig. 15**: Displacement and strain fields obtained with the forward simulation and with the inverse results for all cases shown in **Fig. 14.** The 1$^{st}$ and 2$^{nd}$ columns correspond to the x and y displacement components, respectively. The 3$^{rd}$, 4$^{th}$, 5$^{th}$ columns correspond to three in-plane strain components.

## 3.2 Examples with experimental data

After validating the inverse solver using different simulated cases, we applied the technique on displacement fields measured in the nuclei of two sample cardiomyocytes – one in physiological and another in pathological condition. In solving the inverse problem, the bound of the shear modulus ratio was set to [1, 30]. In **Fig.16**, the raw image, the displacement fields, and the strain fields are shown. We observed that the resulting strain components can be higher than 10%. This demonstrates that the cell nucleus undergoes a large deformation (see **Fig.16(d)-(f)**).

Furthermore, in **Fig.16(m)**, the heterochromatin is primarily gathered on the left side of the cardiomyocyte nucleus, consistently with the H domain shown through pixel grayscale intensity (see **Fig.16(l)**). We also observed that the H:E distributions acquired by the pixel grayscale intensity or the adjoint-based inverse solver are completely different from the strain fields. Thus, it is impossible to determine the H:E distributions qualitatively based on the associated strain fields. Note that our approach is not only able to recover the H domain as this can also be done based on pixel grayscale intensity, but our approach can also identify stiffness heterogeneities within the H domain itself. **Table 3** reports the H:E values at three different regions. They remain within the same range order of magnitude as the average H:E value obtained from the pixel grayscale intensity. Additionally, we conducted a comparison of the reconstruction outcomes under varying regularization factors, as illustrated in **Fig. 17**. Our observations revealed a noteworthy trend: for smaller regularization factors, the boundary delineating the heterochromatin region exhibits a pronounced sharpening effect. Conversely, as the regularization factor increases, the resultant elasticity map of the nucleus tends to manifest an undesirable degree of smoothing, thus potentially compromising the accuracy of the reconstructed model.

For the other cardiomyocyte sample shown in **Fig.18**, the strain level is higher than what we observed in **Fig.16**. Moreover, the H:E distribution identified by the adjoint-based inverse solver (see **Fig.18(m)**) is different from the one acquired by the pixel grayscale intensity (see **Fig.18 (l)**). More specifically, there is a region of very low stiffness in the center of the nucleus in both H:E distributions but its location and dimension are different in **Figs.18(l)** and **18(m)**. Nevertheless, the H:E ratios were in the same range order as the average H:E ratio of [28] (see **Table 3**). The difference between the H:E ratio in region A and the average H:E ratio of ref [28] is only 4%.

Furthermore, it is noteworthy that the trends observed in the reconstructed results under varying regularization factors, as depicted in **Fig.19**, align closely with those depicted in **Fig.17**.

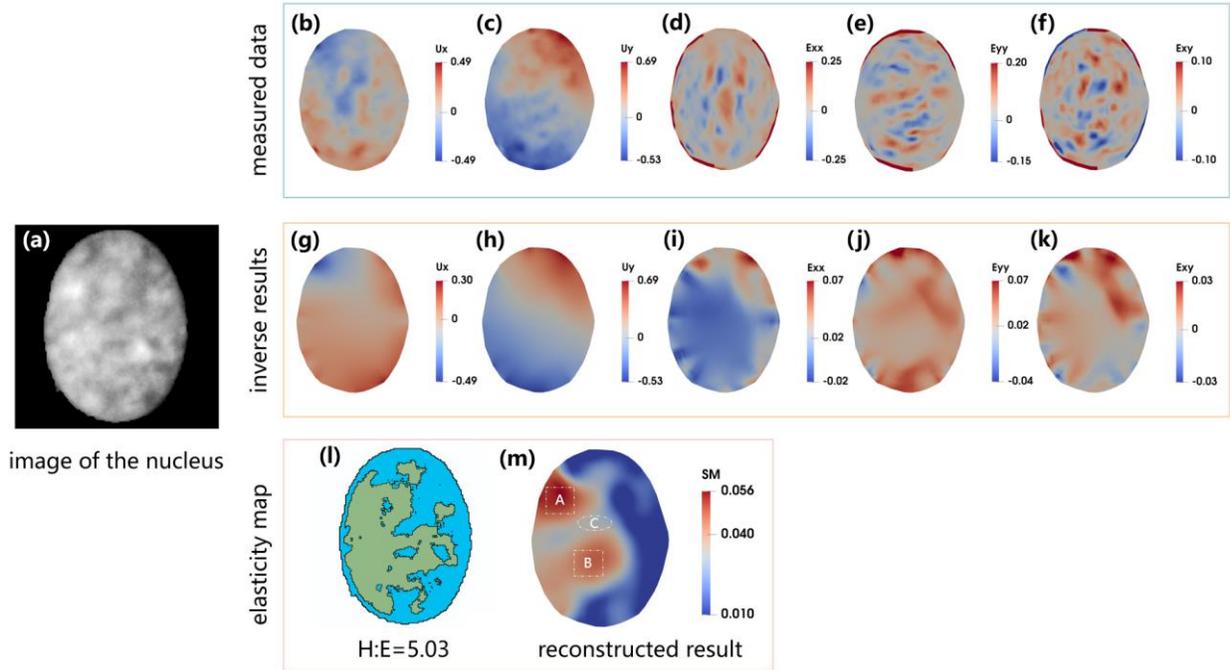

**Fig.16**: Spatial maps of stiffness were computed in the deforming nucleus using inverse methods. (a) Microscopy image of the first nucleus of interest; (b) and (c) Displacement fields measured across the nucleus; (d)-(f) Strain fields computed across the nucleus after differentiation of the displacement fields; (g) and (h) Displacement fields at the last iteration in inverse problem; (i)-(k) Strain fields computed across the nucleus at the last iteration in inverse problem; (l) H and E domains across the nucleus as shown by pixel grayscale intensity [28]; (m) Stiffness map reconstructed with our inverse solver.

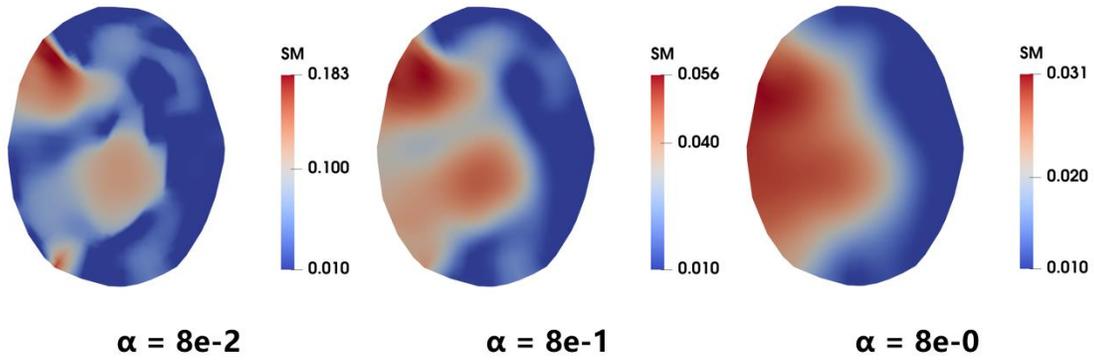

**Fig.17**: Spatial maps of stiffness for varying regularization factors.

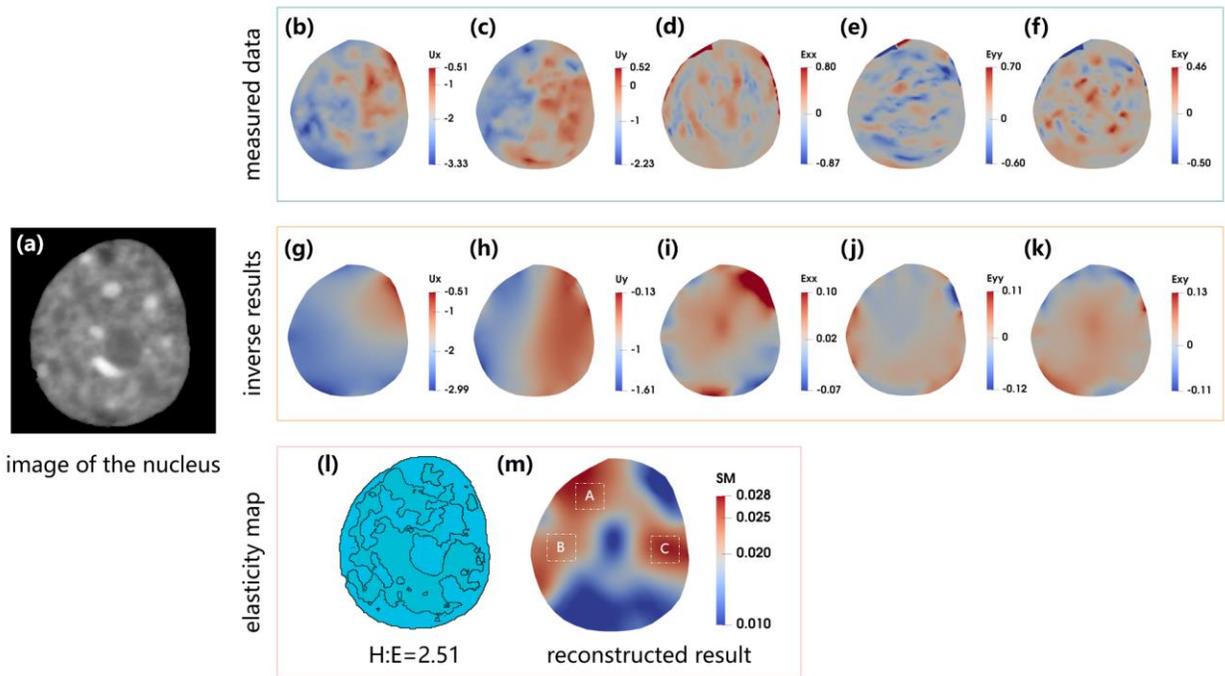

**Fig.18**: Spatial maps of stiffness reveal heterogeneous material properties in the deforming nucleus. (a) Microscopy image of the second nucleus of interest; (b) and (c) Displacement fields measured across the nucleus; (d)-(f) Strain fields computed across the nucleus after

differentiation of the displacement fields; (g) and (h) Displacement fields at the last iteration in inverse problem; (i)-(k) Strain fields computed across the nucleus at the last iteration in inverse problem; (l) H and E domains across the nucleus as shown by pixel grayscale intensity [28]; (m) Stiffness map reconstructed with our inverse solver.

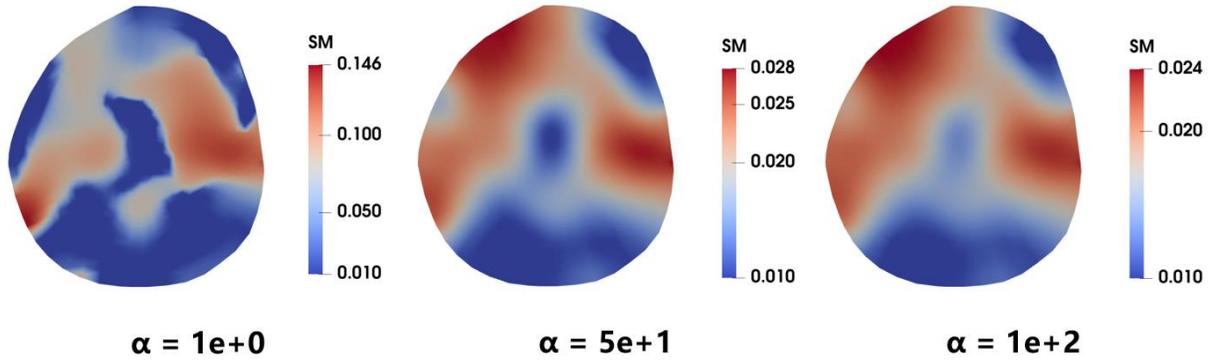

**Fig.19**: Spatial maps of stiffness for varying regularization factors.

**Table 3**: Estimated H:E ratios at regions A, B and C for the two nuclei shown in **Figs**.**16** and **18**(Note that H:E ratio predicted in [28] was solved by an optimization scheme assuming H and E domains are homogeneous)

|  | H:E=5.03[28] | | | H:E=2.51[28] | | |
| --- | --- | --- | --- | --- | --- | --- |
|  | A | B | C | A | B | C |
| H:E | 5.41 | 4.71 | 3.33 | 2.60 | 2.32 | 2.67 |

**4.Discussion**

This paper presents a novel approach to identify the stiffness distribution within the cell nucleus from high-resolution displacement map. The approach relies on an in-house adjoint-based inverse solver. Since the cell nucleus undergoes deformations with strains larger than 5%, the inverse solver is based on the finite deformation theory. The major achievement is to reconstruct heterogeneous stiffness distributions without any *a priori* assumption about the distribution of heterochromatin and euchromatin. Results show that the stiffness distribution in the heterochromatin domain is not homogenous, which can be attributed to the variations in density across the cell nucleus. Another important observation is that the boundaries of the heterochromatin domain cannot be detected from the strain fields due to the amplification of the noise level when computing strain by differentiating the measured displacement fields. The inverse solver addresses these effects by the regularization term. Another important point is that the shear modulus distribution can only be mapped up to a multiplicative factor as only displacements are measured but no force value is available. To obtain quantitative information on the shear moduli of the cell nucleus, tractions at the boundary would need to be measured, or alternatively, if the density of the cell nucleus is known, its spatial variation of shear moduli could be quantified by measuring dynamic deformations [40-41].

The stiffness map of the cell nucleus exhibits significant sensitivity to image quality, thereby presenting an opportunity to enhance the resolution of the stiffness map through improvements in deformation microscopy resolution. **Figs.16** and **18** demonstrate that the adjoint-based inverse solver effectively captures the topology of the heterochromatin region, akin to the information obtained from pixel grayscale intensity. Consequently, employing the pixel grayscale intensity distribution as the initial estimation in the optimization problem holds potential for enhancing both convergence performance and accuracy.

Although the cell nucleus may typically be assumed to be incompressible or nearly incompressible as most living matter [42], experiments using micropipette aspiration have shown significant volumetric strain in the nucleus [39]. Therefore, it was important to investigate the impact of the Poisson's ratio on the identification of nucleus stiffness. **Figs.8** and **10** show that the identified stiffness maps are in agreement with the shape of the heterochromatin domain independently of the Poisson's ratio. Moreover, the estimated H:E stiffness ratio is only slightly influenced by the Poisson's ratio. This observation is consistent with a previous study at the tissue scale [43].

In this study, we employ the assumption of plane stress. Given the typically spherical or oval shape of the nucleus, as referenced in [44], the application of either 2D plane stress or plane strain assumptions may be inappropriate. This could potentially induce a bias in the identification process. For a symmetric model as shown in **Fig.12,** the error generated by the plane stress assumption is notably minor, the identification bias could escalate with variations in shape or loadings such as the case presented in **Fig.14**. Consequently, to overcome this limitation, the implementation of three-dimensional full-field measurements becomes essential in the future.

Our future work will primarily focus on improving the resolution of deformation microscopy and the accuracy of the inverse scheme to increase the robustness and reliability of the proposed method. In this study, the nucleus was assumed to follow the simple neo-Hookean model, and further research is needed to investigate whether the nonlinear elastic behavior of the cell nucleus follows a more complex hyperelastic model under very large deformations. Furthermore, understanding the relationship between the nuclear stiffness distribution and distribution of epigenetic modifiers [45] is crucial because a denser chromatin region is most likely enriched with gene repression specific epigenetic modification. More investigation is needed to relate nuclear stiffness and gene expressions to potentially uncover new insights into the role of nuclear mechanics in cellular biology.

## 5. Conclusion

In this paper, we combined deformation microscopy-based displacement map and an adjoint-based inverse solver to nondestructively identify the nonhomogeneous stiffness distribution of the cell nucleus. The feasibility of the proposed method was successfully demonstrated using simulated data generated by finite-element simulations. Our results indicate that the distribution of the heterochromatin domain can be characterized by the inverse solver, with a maximum relative error of less than 5% for the H:E ratio. We also found that the reconstructed heterochromatin domain is not sensitive to the Poisson's ratio. Finally, we obtained the nuclear stiffness distribution in beating cardiomyocytes using deformation microscopy-based displacement map and previously found two-domain elastography map. Overall, the proposed approach shows great potential for nuclear elastography, with promising directions to investigate mechano-genetics and mechano-epigenetics.

**Declaration of Competing Interest**

The authors declare that they have no known competing financial interests or personal relationships that could have appeared to influence the work reported in this paper.

**Acknowledgments**

The authors acknowledge the support from the National Natural Science Foundation of China (12002075) and the National Natural Science Foundation of Liaoning Province in China (2021-MS-128). The authors additionally acknowledge funding from the National Science Foundation (CMMI 2212121 and CMMI/CAREER 1349735).

# References

[1] Jacobs C R, Huang H, Kwon R Y. Introduction to cell mechanics and mechanobiology. Garland Science, 2012.

[2] Ramérez-Mena A, Andrés-León E, Jesús'Avarez-Cubero M, et al. Explainable artificial intelligence to predict and identify prostate cancer tissue by gene expression. Computer Methods and Programs in Biomedicine, 2023: 107719.

[3] Weijer C J. Collective cell migration in development. Journal of Cell Science, 2009, 122(18): 3215-3223.

[4] Jain S, Cachoux V M L, Narayana G H N S, et al. The role of single-cell mechanical behaviour and polarity in driving collective cell migration. Nature Physics, 2020, 16(7): 802-809.

[5] Lautenschläger F, Paschke S, Schinkinger S, et al. The regulatory role of cell mechanics for migration of differentiating myeloid cells. Proceedings of the National Academy of Sciences, 2009, 106(37): 15696-15701.

[6] Titushkin I, Cho M. Modulation of cellular mechanics during osteogenic differentiation of human mesenchymal stem cells. Biophysical Journal, 2007, 93(10): 3693-3702.

[7] Bongiorno T, Kazlow J, Mezencev R, et al. Mechanical stiffness as an improved single-cell indicator of osteoblastic human mesenchymal stem cell differentiation. Journal of Biomechanics, 2014, 47(9): 2197-2204.

[8] Kliche K, Jeggle P, Pavenstädt H, et al. Role of cellular mechanics in the function and life span of vascular endothelium. Pflügers Archiv-European Journal of Physiology, 2011, 462: 209-217.

[9] Seelbinder B, Ghosh S, Schneider S E, et al. Nuclear deformation guides chromatin

reorganization in cardiac development and disease. Nature Biomedical Engineering, 2021, 5(12): 1500-1516.

[10] Athanasiou K A, Shieh A C. Biomechanics of single chondrocytes and osteoarthritis. Critical Reviews™ in Biomedical Engineering, 2002, 30(4-6).

[11] Huang Y, Tan Y, Tian X, et al. Three-dimensional surgical planning and clinical evaluation of the efficacy of distal tibial tuberosity high tibial osteotomy in obese patients with varus knee osteoarthritis. Computer Methods and Programs in Biomedicine, 2022, 213: 106502.

[12] McCreery K P, Xu X, Scott A K, et al. Nuclear stiffness decreases with disruption of the extracellular matrix in living tissues. Small, 2021, 17(6): 2006699.

[13] Carrasco-Mantis A, Alarcón T, Sanz-Herrera J A. An in silico study on the influence of extracellular matrix mechanics on vasculogenesis. Computer Methods and Programs in Biomedicine, 2023, 231: 107369.

[14] Mora M T, Zaza A, Trenor B. Insights from an electro-mechanical heart failure cell model: Role of SERCA enhancement on arrhythmogenesis and myocyte contraction. Computer Methods and Programs in Biomedicine, 2023, 230: 107350.

[15] Rahman M M, Watton P N, Neu C P, et al. A chemo-mechano-biological modeling framework for cartilage evolving in health, disease, injury, and treatment. Computer Methods and Programs in Biomedicine, 2023, 231: 107419.

[16] Moeendarbary E, Harris A R. Cell mechanics: principles, practices, and prospects. Wiley Interdisciplinary Reviews: Systems Biology and Medicine, 2014, 6(5): 371-388.

[17] Haase K, Pelling A E. Investigating cell mechanics with atomic force microscopy. Journal of The Royal Society Interface, 2015, 12(104): 20140970.

[18] Li M, Dang D, Liu L, et al. Atomic force microscopy in characterizing cell mechanics for

biomedical applications: A review. IEEE Transactions on NanoBioscience, 2017, 16(6): 523-540.

[19] Wirtz D. Particle-tracking microrheology of living cells: principles and applications. Annual Review of Biophysics, 2009, 38: 301-326.

[20] Wu P H, Aroush D R B, Asnacios A, et al. A comparison of methods to assess cell mechanical properties. Nature Methods, 2018, 15: 491-498.

[21] Grasland-Mongrain P, Zorgani A, Nakagawa S, et al. Ultrafast imaging of cell elasticity with optical microelastography. Proceedings of the National Academy of Sciences, 2018, 115(5): 861-866.

[22] Graff K F. Wave motion in elastic solids. Courier Corporation, 2012.

[23] Sinkus R, Lorenzen J, Schrader D, et al. High-resolution tensor MR elastography for breast tumour detection. Physics in Medicine & Biology, 2000, 45(6): 1649.

[24] Papazoglou S, Hamhaber U, Braun J, et al. Algebraic Helmholtz inversion in planar magnetic resonance elastography. Physics in Medicine & Biology, 2008, 53(12): 3147.

[25] Baumann K. A vision of 3D chromatin organization. Nature Reviews Molecular Cell Biology, 2017, 18(9): 532-532.

[26] Maloney J M, Nikova D, Lautenschläger F, et al. Mesenchymal stem cell mechanics from the attached to the suspended state. Biophysical Journal, 2010, 99(8): 2479-2487.

[27] Grewal S I S, Jia S. Heterochromatin revisited. Nature Reviews Genetics, 2007, 8(1): 35-46.

[28] Ghosh S, Cuevas V C, Seelbinder B, et al. Image-Based Elastography of Heterochromatin and Euchromatin Domains in the Deforming Cell Nucleus. Small, 2021, 17(5): 2006109.

[29] Alisafaei F, Jokhun D S, Shivashankar G V, et al. Regulation of nuclear architecture, mechanics, and nucleocytoplasmic shuttling of epigenetic factors by cell geometric

constraints. Proceedings of the National Academy of Sciences, 2019, 116(27): 13200-13209.

[30] Ghosh S, Seelbinder B, Henderson J T, et al. Deformation microscopy for dynamic intracellular and intranuclear mapping of mechanics with high spatiotemporal resolution. Cell Reports, 2019, 27(5): 1607-1620. e4.

[31] Lomakin A J, Cattin C J, Cuvelier D, et al. The nucleus acts as a ruler tailoring cell responses to spatial constraints. Science, 2020, 370(6514): eaba2894.

[32] Venturini V, Pezzano F, Catala Castro F, et al. The nucleus measures shape changes for cellular proprioception to control dynamic cell behavior. Science, 2020, 370(6514): eaba2644.

[33] Hansen P C. Analysis of discrete ill-posed problems by means of the L-curve. SIAM Review, 1992, 34(4): 561-580.

[34] Hughes T J R, Franca L P, Balestra M. A new finite element formulation for computational fluid dynamics: V. Circumventing the Babuška-Brezzi condition: A stable Petrov-Galerkin formulation of the Stokes problem accommodating equal-order interpolations. Computer Methods in Applied Mechanics and Engineering, 1986, 59(1): 85-99.

[35] Oberai A A, Gokhale N H, Feijóo G R. Solution of inverse problems in elasticity imaging using the adjoint method. Inverse Problems, 2003, 19(2): 297.

[36] Mei Y, Goenezen S. Quantifying the anisotropic linear elastic behavior of solids. International Journal of Mechanical Sciences, 2019, 163: 105131.

[37] Mei Y, Goenezen S. Mapping the viscoelastic behavior of soft solids from time harmonic motion. Journal of Applied Mechanics, 2018, 85(4): 041003.

[38] Byrd R H, Lu P, Nocedal J, et al. A limited memory algorithm for bound constrained optimization. SIAM Journal on Scientific Computing, 1995, 16(5): 1190-1208.

[39] Rowat A C, Lammerding J, Ipsen J H. Mechanical properties of the cell nucleus and the effect of emerin deficiency. Biophysical Journal, 2006, 91(12): 4649-4664.

[40] Wijesinghe P, Johansen N J, Curatolo A, et al. Ultrahigh-resolution optical coherence elastography images cellular-scale stiffness of mouse aorta. Biophysical Journal, 2017, 113(11): 2540-2551.

[41] Leartprapun N, Iyer R R, Untracht G R, et al. Photonic force optical coherence elastography for three-dimensional mechanical microscopy. Nature Communications, 2018, 9(1): 2079.

[42] Chan C J, Li W, Cojoc G, et al. Volume transitions of isolated cell nuclei induced by rapid temperature increase. Biophysical Journal, 2017, 112(6): 1063-1076.

[43] Mei Y, Deng J, Guo X, et al. Introducing regularization into the virtual fields method (VFM) to identify nonhomogeneous elastic property distributions. Computational Mechanics, 2021, 67: 1581-1599.

[44] Webster M, Witkin L.W, Cohen-Fix O. Sizing up the nucleus: nuclear shape, size and nuclear-envelope assembly. Journal of Cell Science, 2009, 122 (10): 1477-1486.

[45] Nims R J, Pferdehirt L, Guilak F. Mechanogenetics: harnessing mechanobiology for cellular engineering. Current Opinion in Biotechnology, 2022, 73: 374-379.